\documentclass[journal]{vgtc}                
\ifpdf
  \pdfoutput=1\relax                   
  \usepackage{graphicx}                
  \DeclareGraphicsExtensions{.pdf,.png,.jpg,.jpeg} 
\else
  \usepackage{graphicx}                
  \DeclareGraphicsExtensions{.eps}     
\fi%

\graphicspath{{figures/}{pictures/}{images/}{./}} 

\usepackage{microtype}                 
\PassOptionsToPackage{warn}{textcomp}  
\usepackage{textcomp}                  
\usepackage{mathptmx}                  
\usepackage{times}                     
\usepackage{cite}                      
\usepackage{tabu}                      
\usepackage{booktabs}                  

\onlineid{0}

\vgtccategory{Research}
\vgtcpapertype{Poster}

\title{A Pilot Study on The Impact of Stereoscopic Display Type on User Interactions Within A Immersive Analytics Environment}

\author{Adam S. Williams, Xiaoyan Zhou, Michel Pahud, and Francisco R. Ortega}
\authorfooter{
\item
 Adam Williams, Xiaoyan Zhou, and Francisco Ortega are with Colorado State University. E-mails: AdamWil@colostate.edu, xiaoyan.zhou@colostate.edu, FOrtega@colostate.edu.
\item
 Michel Pahud is will Microsoft Research. E-mail: mpahud@microsoft.com.
}

\shortauthortitle{Biv \MakeLowercase{\textit{et al.}}: Global Illumination for Fun and Profit}

\abstract{
Immersive Analytics (IA) and consumer adoption of augmented reality (AR) and virtual reality (VR) head-mounted displays (HMDs) are both rapidly growing. When used in conjunction, stereoscopic IA environments can offer improved user understanding and engagement; however, it is unclear how the choice of stereoscopic display impacts user interactions within an IA environment. This paper presents a pilot study that examines the impact of stereoscopic display type on object manipulation and environmental navigation using consumer-available AR and VR displays. This work finds that the display type can impact how users manipulate virtual content, how they navigate the environment, and how able they are to answer questions about the represented data.
} 


\CCScatlist{ 
 \CCScat{H.5.1}{Multimedia Information Systems}%
{Artificial, augmented, and virtual realities}{};
 \CCScat{H.1.2}{User/Machine Systems}{Human factors}{}
}

\teaser{
  \includegraphics[width=1\columnwidth]{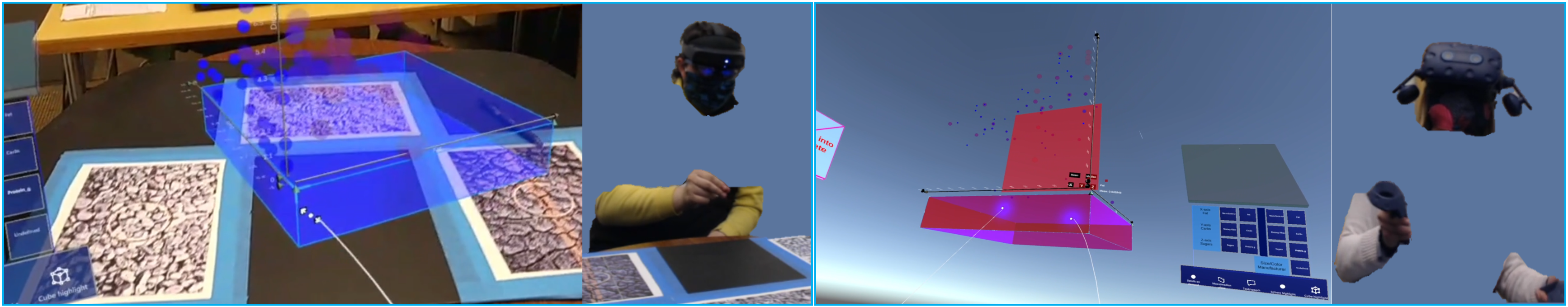}
  \caption{Left: Augmented reality participant, Right: Virtual reality participant}
  \label{fig:teaser}
}

\vgtcinsertpkg

\begin{document}


\firstsection{Introduction}

\maketitle

Augmented reality (AR) and virtual reality (VR) head-mounted displays (HMDs) can offer many benefits over a standard desktop workspace, including, increased immersion, accessible data display, and improved user engagement~\cite[Chapter~1]{IABOOK}. Three-dimensional (3D) data visualization environments, also called immersive analytics (IA) environments, often leverage the benefits of stereoscopic displays in attempts to improve user understanding of the represented data~\cite{MOT+18, DRO+15}. A necessary step towards selecting a stereoscopic display for data analysis tasks is understanding how that choice can impact user interactions and navigation within the rendered IA environments.

Researchers have started to shed light on this issue by examining how users manage virtual content in VR IA environments~\cite{BAT+20}. To date most work in IA has been conducted using VR-HMDs~\cite{FON19}. Less work has examined IA use in stereoscopic AR HMDs or how the choice of stereoscopic display impacts user interactions in IA environments. Works that have compared across display types often do so at a granular level; comparing object manipulations alone~\cite{KRI+18}, or visualization understanding~\cite{WHI+20}.

To leverage prior work in IA that was done using VR-HMDs when developing for AR-HMDs, a more in-depth understanding of how interactions differ between AR and VR displays needs to be established. This understanding can improve both the development of IA systems and the development of interaction techniques for these environments. Towards that goal, this preliminary work builds on prior works by observing how people interact with, navigate, and manage virtual content in a single IA environment across two types of stereoscopic displays with limited constraints put on user interactions or navigation. The goal of this pilot study is to open conversations around AR compared to VR IA use. The authors hope that this work generates a starting point for future larger and more controlled AR VR comparisons.

\section{Methods}

\begin{figure*}[!htb]
  \centering
  \includegraphics[width=2\columnwidth, keepaspectratio]{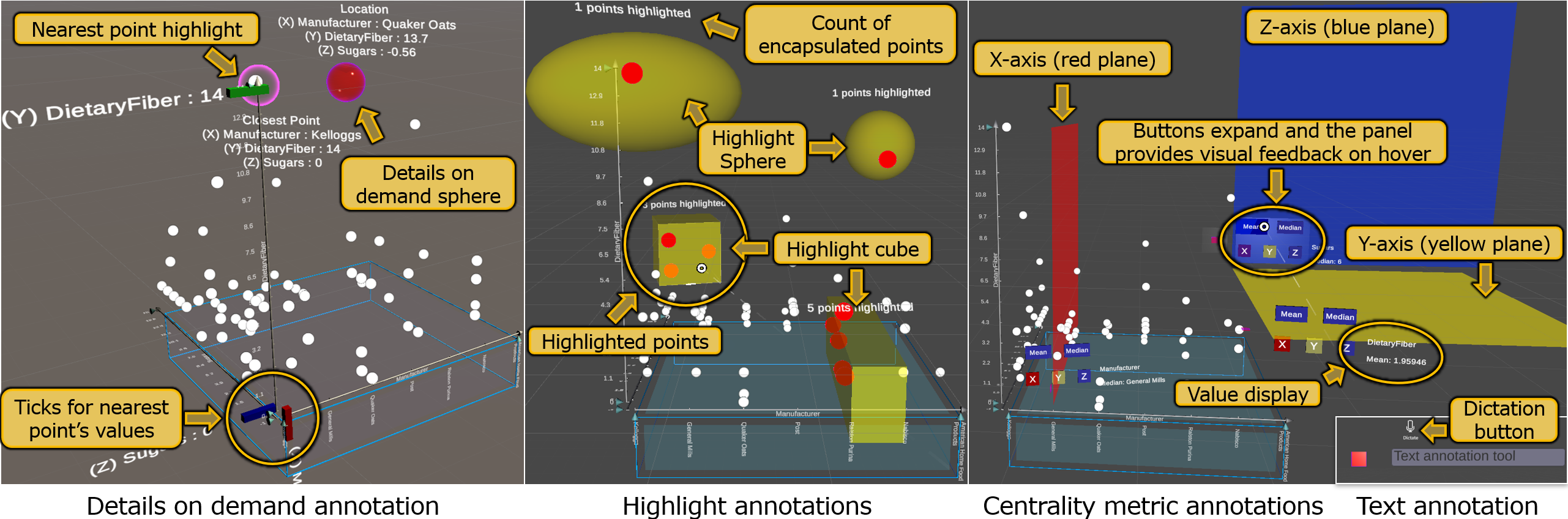}
  \caption{All annotations provided by the system, annotations and annotation features are labeled}
  \label{fig:AllAnnotations}
\end{figure*}

\begin{figure}[!htb]
  \centering
  \includegraphics[width=.75\columnwidth, keepaspectratio]{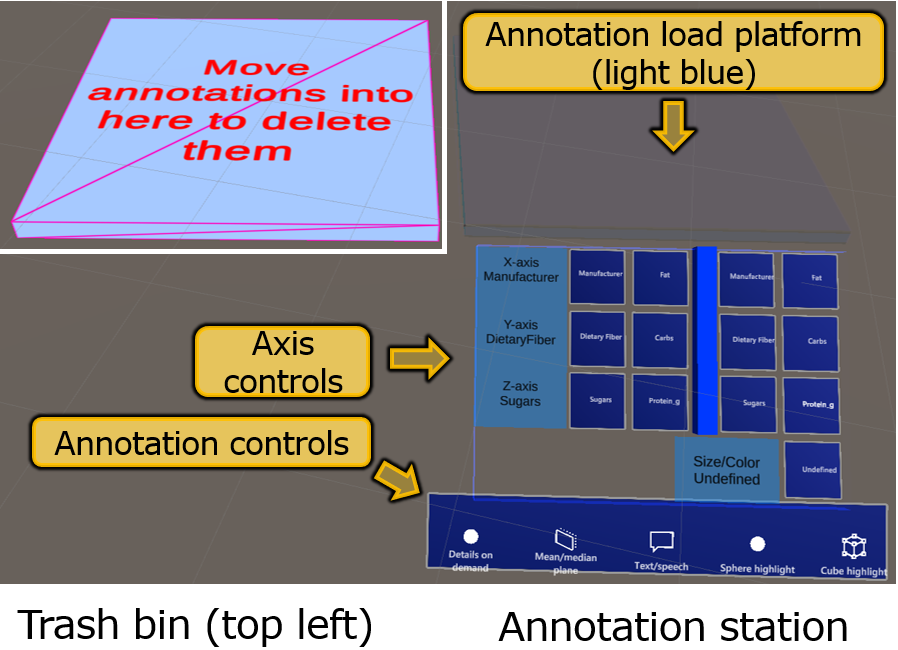}
  \caption{Trash bin and annotation station with labels}
  \label{fig:controls}
\end{figure}

Twelve volunteers participated in this between-subjects AR-VR IA environment interaction comparison study. These volunteers were randomly split into two groups of 6. Each group was assigned one device, either a VR-HMD or an AR-HMD. During this experiment, participants used 5 types of annotations while interpreting a 3D scatterplot featuring a cereal nutrition dataset (Figure~\ref{fig:AllAnnotations}). A trash bin and annotation station were included in the environment to allow users to delete unwanted annotations, generate new annotations, and change the visualization's axis mappings (Figure~\ref{fig:controls}).

\subsection{Experiment Design}

Participants would arrive at the lab and were asked to sit at a round table (Figure~\ref{fig:teaser}). Participants were first given informed consent, a demographics survey, the short graph literacy scale (SGLS), and the VZ-2 paper folding test. After completing the pre-study forms, participants were told about the experiment at a high level and donned either an AR or VR HMD. The SGLS consists of 4 questions that can be used to assess graph literacy and understanding~\cite{OKA+19}. The VZ-2 paper folding test assesses visual and spatial reasoning skills by having participants determine where holes would appear on a hypothetical unfolded piece of paper when a pencil is pushed through a folded piece of paper~\cite{EKS+76}.

Participants then completed a training session that covered loading, placing, and adjusting the visualization (i.e., the scatter-plot), trash tool, and annotation/visualization control station (Figure~\ref{fig:controls}). Participants were asked to place these objects where they would be comfortable interacting with them, acknowledging that object placement could be changed at any time. After the participant's arranged their work space, they were told how to change the dimension mappings on the visualization (i.e., changing the x-axis) and the five provided annotation tools were explained. These tools were details on demand (DoD), the mean/median plane (centrality plane), a text entry box, and two highlight volumes (Figure~\ref{fig:AllAnnotations}). All dimension mapping changes and annotation tool generations were accomplished using the provided annotation/visualization control station (Figure~\ref{fig:controls}).

After training participants were instructed to navigate the visualization, interact with the tools, and generate questions about the data that could be asked to other users. This process (phase one) lasted 15 minutes or until the participant asked for the next phase to be started. Phase two consisted of a 15-minute session where the researcher would ask questions about the dataset. Phase two ended once a participant answered all questions, when 15 minutes had elapsed, or when a participant requested to end the session. Some examples of the questions asked during phase two are ``What manufacturer has the fewest cereals represented?'', ``What is the highest sugar content contained in the scatter-plot'', and ``Which manufacturer or manufacturers have the largest portion of their cereals containing lower than average fat''.

\begin{figure}[htb]
  \centering
  \includegraphics[width=.85\columnwidth, keepaspectratio]{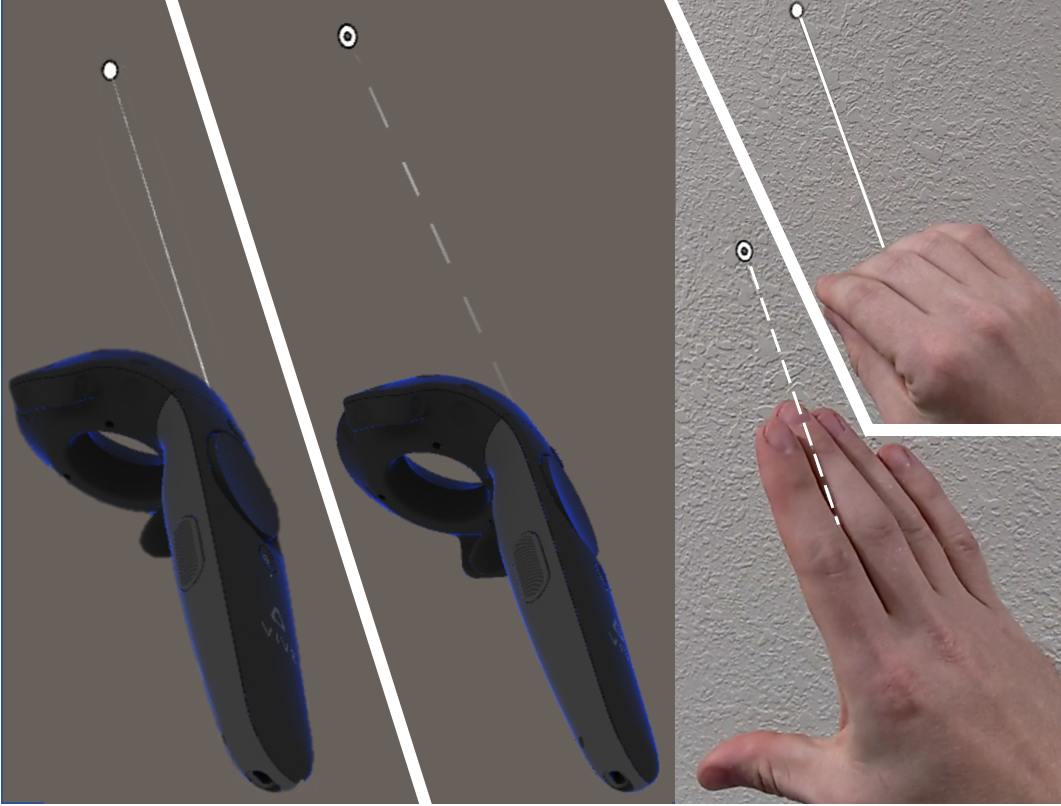}
  \caption{Left: VR selected and hover ray-casts, Right: AR hover and selected ray-casts}
  \label{fig:rayCasts}
\end{figure}

In both groups, a ray-cast technique was used to interact with virtual content. In VR the Vive controller was used to move the ray-cast and selection was achieved by pressing a button (Figure~\ref{fig:rayCasts}). In AR, this ray-cast extended from one's hand, and selection was triggered by pinching (Figure~\ref{fig:rayCasts}). Only the headset and the ray-cast technique used were different between the two groups. 

\begin{figure}[htb]
  \centering
  \includegraphics[width=.99\columnwidth, keepaspectratio]{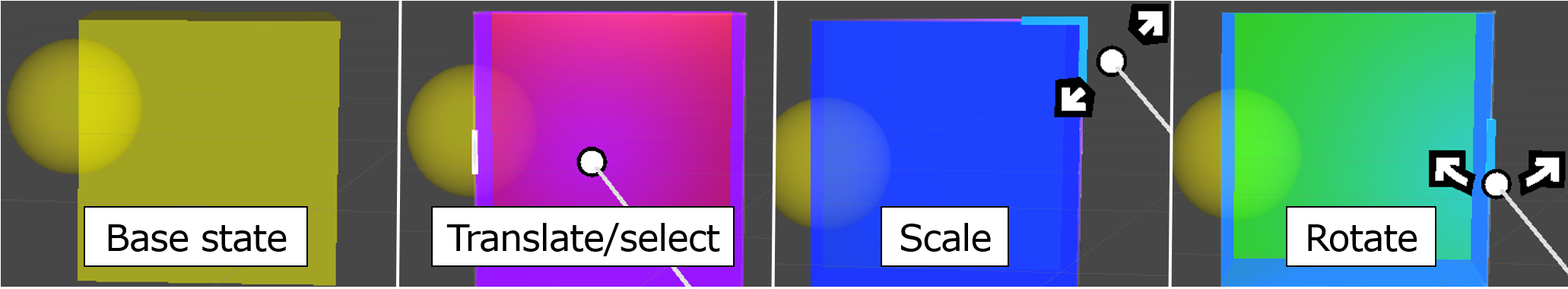}
  \caption{Feedback states used}
  \label{fig:feedback}
\end{figure}

Feedback mechanisms included distinct audio and visual cues for each interaction. As an example, to select and translate an object the user would hover their ray-cast over the object which would trigger a proximity-based color change near the ray-cast's intersection point with the object. Then the selected object would turn red and play a translation-specific audio clip (Figure~\ref{fig:feedback}). When released the object would return to its base color and play a different but similar release audio clip. Similar sequences of events occur for scale and rotation, and virtual button presses (Figure~\ref{fig:feedback}).

The main goal of this study was to observe how people interact with the environment, not to measure the accuracy of responses to questions or to identify individual annotation tool use. Participants that struggled to answer the questions asked during phase two were given easier questions. After the experiment, participants removed the HMD and completed a 0-100 scale NASA Task Load Index (TLX) to measure their perceived workload~\cite{HAR+88}.

\subsection{Participants}

Twelve participants were randomly split into two groups. All participants confirmed that they were comfortable interacting with 2D scatter-plot charts and had normal or corrected to normal vision. The VR group had an average age of 21.5 years with an SD of 3.99 years. Five participants had used an AR headset for 30 minutes or less prior to this experiment. All participants were right-handed. One participant played VR games for 3 hours a week and one played them for 1 hour a week. The VR group consisted of $5$ females with $1$ male. The AR group had an average age of 25 years with an SD of 4.78 years. Three participants had used an AR headset for 30 minutes or less before this experiment. Four were right-handed. Two participants indicated that they play VR games for 1 hour a week. The AR group was composed of $5$ males and $1$ female. This gender imbalance was caused by participant session cancellations and new participant recruitment mid-way through data collection and difficulties recruiting participants during the COVID-19 pandemic.
 
\subsection{Apparatus and Data Collection}

VR sessions used an HTC Vive Eye Pro with a 110 degree field of view~\footnote{https://www.vive.com/us/product/vive-pro/}. The Vive was connected to a Windows 10 computer with 32 GB of RAM, an Intel i9-9900k CPU (3.60 GHz), and an Nvidia GeForce 2080ti. The AR sessions were conducted using a Microsoft Hololens 2 which has 52 degree field of view~\footnote{https://www.immersiv.io/blog/hands-on-hololens-2-review}. The IA system was developed using Unity version 2019.2.18f1, the MRTK version 2.5.1, Vuforia version 9.6.3, and the Immersive Analytics Toolkit (IATK)~\cite{IATK}.

Video was collected using a web camera and recordings from the rendered environment. The system collected log data for all events (e.g., a dimension mapping change). Log data was combined across participants and cleaned using R. Results are reported using box plots where possible. The mean and SD values for data are provided for numeric data. These results are being reported as Observational trends. With the small sample size for each group, significance tests are not reported.

\section{Results}

\subsection{Surveys}

On average the AR group scored higher on the SGLS than the VR group. The scores for the SGLS were 92.86\% (SD 11.29\%) for the AR group and 75\% (SD 14.43\% ) for the VR group. In the AR group, there were top and bottom performers that scored 100\% on both sections of the SGLS. The AR group outperformed the VR group at the SGLS while also having more participants struggle to finish tasks in the environment. This indicates that the SGLS may not provide a clear signal of a participant's ability to navigate stereoscopic IA environments.

VR participants scored slightly higher than AR group for spatial reasoning ability and had lower variability in their scores. The mean score for the paper folding test in the VR group was 74.14\% (SD of 12.39\%) whereas the AR group had 72.50\% (SD 20.36\%). The AR group score variation was caused by two participants receiving low scores, one receiving a 50\% and the other a 40\%. The lowest two scores in the VR group were 55\% and a 65\%. Similar to the SGLS the paper folding task scores were not strong indicators of participant performance.

\subsection{Time Spent in Environment}

\begin{figure*}[!htb]
  \centering
  \includegraphics[width=1.75\columnwidth, keepaspectratio]{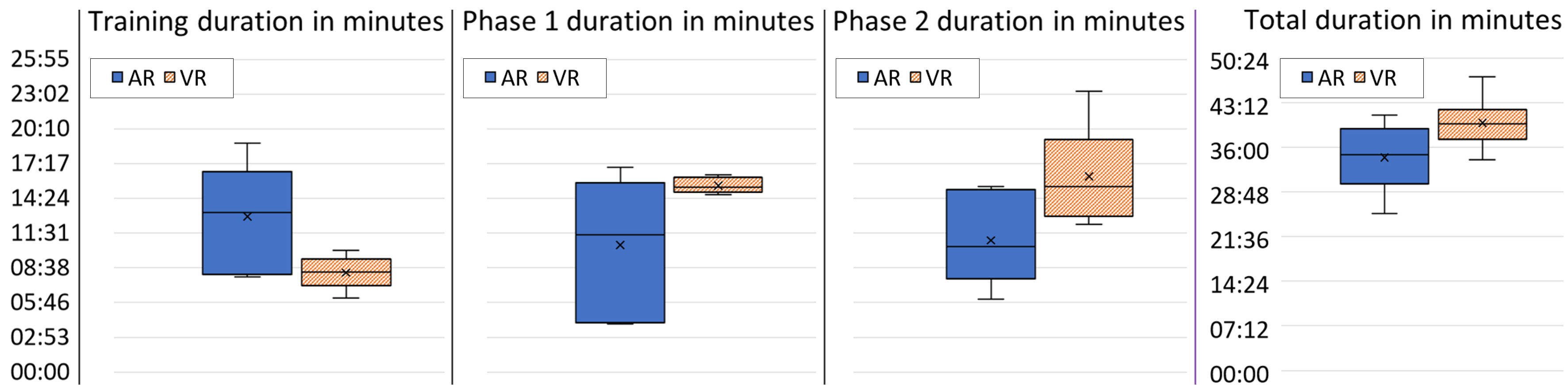}
  \caption{Times that participants spent in different portions of the experiment by device condition}
  \label{fig:SessionTimes}
\end{figure*}

The AR group spent more time training and less time in the rest of the experiment (Figure~\ref{fig:SessionTimes}). The reduced time in the environment was caused by participant request to end phase one and/or phase two early. The three AR group participants that were least able to navigate the environment and answer questions about the data exhibited similar tendencies. These three persons took nearly twice as long as the top three performers to complete the training session (average of 16:43 minutes compared to 9:03) and spent less time both phases, often stopping all interactions with the visualization towards end of each phase. The first portion of the training sessions covered how to use the ``pinch'' gesture interaction for AR users and the controller for VR participants. The pinch gesture took more time to learn than the controller, which could account for some of the differences in training time between the two conditions. All participants had to successfully interact using either the controller or the pinch gesture enough times to place the visualization, annotation station, trash bin, and one of each annotation prior to ending the training session.

The VR group had less deviation in their times and did not have anyone request to end early. Participants in the VR group finished the training phase in 8:14 minutes on average where the AR group took 12:53 minutes on average. Interestingly two of the VR group participants chose to stay in the environment for longer than 15 minutes during the second phase, staying instead for 23:18 and 17:56 minutes. 

\subsection{Experimental Tasks}

AR participants were much more likely to stop interacting with the visualization during phase 1, seen in the three least performant participants. Only two participants in either device group generated questions. In the AR group these participants generated 9 and 4 questions where in the VR group they asked 1 and 4 questions. 

During phase two AR participants answered an average of 5.83 questions of 11 total (SD 2.67) where in VR participants answered an average of 9.5 questions (SD 1.26). The top three performers from the AR group in isolation answered an average of 8.33 questions (SD 1.25). The three less preferment AR participants answered 3, 3, and 4 questions each. In VR all participants continued to answer questions until the end of the session or until 11 questions were asked. One participant in the VR group chose to stay in the environment an extra 8:18 minutes to answer the final question asked.

\subsection{Participant Interactions}

VR participants tended to interact with further away visualizations and held their hands closer to their body. The AR group interacted with a closer visualization. With the visualization being within reach, AR participants often held their hands near the visualization and thus further away from their bodies. Both groups struggled to complete interactions using the ray-cast selection technique. In both groups ray-cast movement during selection and natural jitters in arm movement caused inaccuracies in selection. 

\subsubsection{Visualization Size}

\begin{figure*}[!htb]
  \centering
  \includegraphics[width=1.75\columnwidth, keepaspectratio]{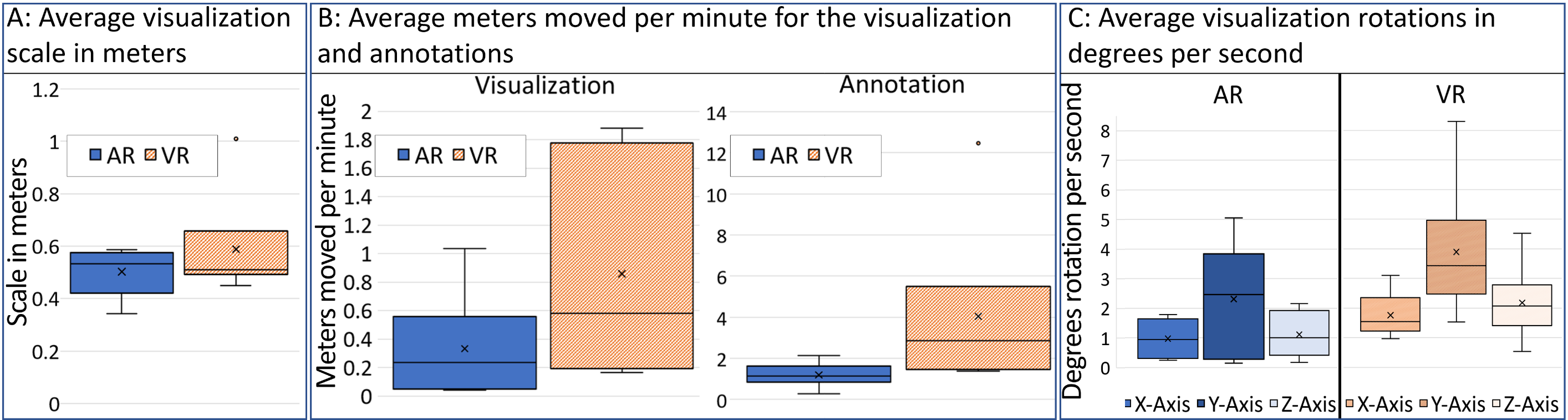}
  \caption{Box-plots from left to right: Average visualization scale (A), Average meters moved (B), Average rotations made (C)}
  \label{fig:multiGraph}
\end{figure*}

The visualization always began as a .42 X .42 X .42 meter cube. Participants were able to resize the visualization at any time. Most participants resized the visualization a few times at the beginning of a session and then left it alone. The VR group used larger scatter plots than the AR group with an average visualization size of .588 meters (SD .191) compared to .502 meters (SD .085) (Figure~\ref{fig:multiGraph}~A). The largest visualization was in the VR group, with a scale of 1.01 meters. That participant also placed their visualization further away than other participants. When asked about the visualization size after the session the participant did not believe that the visualization was scaled to 1 meter, possibly due to the combination of it being both larger and further away. People across the VR condition tended to sit further back from the visualization and keep their hands closer to their bodies. The AR condition could see the desk in front of them and all placed the visualization on that desk. No participants commented on the field of view being a contributing factor to their interactions in the environment.

\subsubsection{Visualization and Annotation Movement}

On average the VR group moved the visualization more per minute of the experiment than the AR group (Figure~\ref{fig:multiGraph}~B). The VR group moved the visualization an average of .859 meters per minute (SD .716) where the AR group only moved the visualization .334 meters per minute (SD .344). Annotations were also moved more by the VR group with an average movement of 4.05 meters per minute (SD 3.832) and 1.198 meters per minute (SD .55) respectively. 

Only three participants moved the annotation station from where it was loaded on their left to their right, all in the VR group. VR participants were also more likely to place the annotation station above their shoulders. The AR group participants placed the annotation station near the surface of the physical table in-front of them.

\subsubsection{Visualization Rotation}

In this environment an x-axis rotation is pitch, a y-axis rotation is yaw, and a z-axis rotation is roll. Rotations about the x-axis were the least performed rotation with a mean of .97 (SD .71) degrees/second (Deg/Sec) for the AR group and 1.76 Deg/Sec (SD .76) for the VR group (Figure~\ref{fig:multiGraph}~C). Rotations about the z-axis were the next least used rotation at 1.11 Deg/Sec (SD .78) for the AR group and 2.18 Deg/Min (SD 1.30) for the VR group. The most performed rotation was yaw or rotating the visualization about the y-axis. In AR participants performed slightly fewer yaw rotations with an average of 2.31 Deg/Sec (SD 1.92) compared to 3.89 Deg/Sec (SD 2.32) for the VR group. In addition to the differences in rotations, VR participants were observed moving their head and upper body to view the data from different angles more frequently than participants in the AR condition.


\subsection{NASA TLX}

In general, there were limited differences between AR and VR conditions for frustration (38.33 AR vs 36.67 VR) and overall workload (55.42 AR vs 52.08 VR). These low scores for frustration and overall workload are interesting when considering the differences in interaction techniques between the two devices. These scores imply that the selection technique (i.e., button vs pinch) and the ray-cast movement type (i.e., controller vs hand) did not contribute to widely varied frustration scores. The physical demand was more varied and slightly higher for the AR group than the VR group (mean of 46.67 AR vs 30.83 VR). This could be excepted as VR controllers can be used with less movement than mid-air gestures. VR participants perceived that they were using more mental and total effort than the AR group (VR mean 81.67 SD 14.63, AR mean 59.17 SD 18.12). This difference might have been contributed to by the difference in engagement between the two groups where VR group interacted with the environment more fully and for longer than the AR group.

\subsection{Other Trends}

The VR participants had a difficult time determining where the z-axis points were. One VR group participant used a highlight annotation which also displayed the count of highlighted points to check their own count of points in that same area (Figure~\ref{fig:AllAnnotations}). The AR group could use real-world visual cues to determine the relative positions of things in the environment, helping mitigate depth estimation issues. Instead, the AR group had more concerns regarding the visibility of color mappings.

\section{Discussion}

This paper has taken the approach of not focusing on the correctness of interactions, questions, or answers. Instead, this work focused on how participants navigated and interacted with the environment. With limited prior work examining how stereoscopic display type impacts user interactions in the same IA environment these findings can help inspire future IA use, research, and development.

We believe that the combination of a complex environment, the optical see-through stereoscopic display, and mid-air gesture interactions were the main contributors to the struggles of the bottom performers of the AR group. Two of the top performers in the AR group had experience using VR-HMDs, meaning that they had used ray-cast interactions before. This base level of familiarity with ray-casting may have allowed these experienced users to focus more on the data rather than navigating the environment itself.


\subsection{Time in Environment}

VR participants spent more time on average in the environment than the AR group. This was also true when comparing the VR group to the top AR performers and the bottom AR performers separately. VR participants also completed the training more quickly than AR participants. These training times reflect the amount of time it took the participant to interact with each tool in the system, suggesting that VR participants picked up tools and features of this environment more quickly than AR participants. 

Increased time spent in the environment may also be related to the immersion that VR participants felt. VR users could not see the outside world, only the virtual environment, causing them to focus more on the tasks given. This additional focus may be reflected in the differences between the AR and VR NASA TLX mental effort scores. It is unclear why AR participants interacted with the system less, even among the participants that were skilled at using the ray-casts. It might be that seeing the real world kept them from getting fully immersed in interactions with the environment.

\subsection{Surveys Used}

The surveys used were selected because they measure skills relevant to this task (i.e., spatial reasoning, graph literacy) and in the case of the paper folding test, have been used by prior work in IA~\cite{MAC+19, LAG+18, WAG+21}; however, the results of these surveys did not provide a clear signal on participant ability to interact with this environment. The participants that did poorly on the SGLS or the paper folding test were not more likely to perform poorly in this environment. It is possible that high 2D graph literacy does not entirely transfer to 3D graph literacy. The additional dimensions of data displayed may require a different form of graph literacy. In this environment, the visualization was directly manipulable. The direct interaction with 3D objects helped participants who self-reported low spatial reasoning (corroborated by the paper folding test results) to perform well in this environment. One participant who self-reported having aphestatia (i.e., a condition where they are unable to summon mental imagery on demand) noted that interacting with the 3D visualization made understanding the data easier for them since they did not have to attempt to mentally compare different 2D graph states.

\subsection{IA Experiment Design Guidelines}

Introducing people to this IA environment utilizing stereoscopic displays was difficult. Participants needed interactive training sessions and live interactions with the system before they were able to perform the experimental tasks. Even by the end of the sessions, participants often commented that the system was unfamiliar to them. IA researchers should plan on performing multiple sessions with participants. Ideally, there should be an initial training session where participants become familiar with the environment and interactions used. At a later point participants could return to complete the experiment. Using this design, researchers could observe how quickly the interaction techniques are remembered by users, providing insights on any differences in retention between the devices. If multiple sessions are not an option, recruiting for prior VR experience may be beneficial as it could help participants more quickly acclimate to the environments and interactions used. 

The participants in this study gained a better understanding of the system when they were actively interacting with the system during training, suggesting that researchers might want to avoid video-based instructions. Moreover, VR-HMDs may be better suited for training. The VR participants were more engaged with the system, interacting more with each tool and with the visualization. This is seen in the lower training times and increased performance of the VR participants. Training participants on the system in VR can tap into that engagement and help reduce the difficulty of learning the system for new users.

\subsection{Observations}

Over the experimental sessions, several interesting themes in user behaviors were noted. Participants using the VR-HMD typically set a larger visualization and placed it without regard to the real-world, often placing it further away from themselves than the AR group. That placement resulted in VR participants interacting with the visualization from a greater distance than AR users (Figure~\ref{fig:teaser}). In AR participants would place the visualization on the desk in front of them. Once placed, participants would interact with the visualization closely, often holding their hands near the visualization (Figure~\ref{fig:teaser}). Researchers could leverage that placement strategy to incorporate passive haptic feedback into the table where the users sit when interacting with an AR IA system. 

Apart from the visualization's scale and placement differences, there were differences in how VR users managed and navigated their virtual space. One such difference is that members of the VR group were the only ones who moved the annotation station from their left, where it was generated, to their right. With all VR participants and 4 of the 6 AR participants being right-handed, it was interesting that only a few participants in the VR condition chose to move the system's most interacted with the tool to their dominant side. Additionally, and opposite to our original expectations, VR users were more likely to move around to view the data from different angles. These users did not walk around, but they did stand, lean, and move their upper bodies. This was in contrast to the AR users who were more likely to rotate or move the visualization. 

\section{Limitations and Future Work}

The gender imbalance may have impacted the results causing the differences to be between males and females and not between AR and VR. More work is needed to tell the level of impact that the gender imbalance had on this study. The SGLS and paper folding test scores were not strongly related to participant performance. The sample size used in this study makes the survey's actual correlation with participant performance uncertain; however, it would be worthwhile for IA researchers to develop a 3D graph literacy scale that can be completed in a stereoscopic HMD. It seems likely that 2D graph literacy and 3D graph literacy may be related but different skill sets.

This work shows trends that they may be differences between AR and VR IA system use. To further hone in on the extent of these differences future work can isolate and expand upon specific components of this work. Using a headset that can be set to a video-pass through mode and a VR mode would eliminate confounds caused by seeing the real world and confounds caused by differences in the stereoscopic displays themselves. Holding the interaction techniques constant would mitigate the impacts of hand vs controller-based ray-casting. The use of a mid-air pen is one possible solution for holding the interaction technique constant between devices. 

This work did not directly examine the effect of the device's field of view on participant interactions. Some of the observed differences may have been influenced by these differences in field of view. Future work could force the HMD with the larger field of view to only show the range that the other HMD was capable of displaying. 

Several questions have come up due to this work. A more in-depth examination of the differences in time spent in the environment and observed levels of participant engagement between the two devices would help better guide future device selection for IA use. Another interesting question is about the use of space and the impact of distractions from the real world. VR participants placing the visualization further away from themselves while AR participants placed it in front of them may have been caused by the AR participants seeing the real-world desk. The AR participants reduced time spent in the environment or their lower levels of engagement might have also been influenced by their seeing the real world.

\section{Conclusion}

This study is one of the first studies in IA to compare participant interactions and navigation between AR and VR HMDs using the same virtual environment. This study found that not all participants in AR were able to interact successfully in the system, potentially causing those participants to perform poorly and spend less time in the environment. These difficulties may have stemmed from struggles in understanding how to select and navigate content in the environment and a lessened sense of immersion caused by seeing the real world. With those difficulties encountered early on, these AR participants became disengaged with the system, interacting less with it, and answering fewer questions about it. AR participants also spent longer in the training phase but less time in the other phases of the experiment.

There were also differences in how participants in AR compared to those in VR navigated, interacted with, arranged, and understood virtual content. In VR participants were more immersed in the environment, leading to the increased time spent in the system, more interactions with the virtual content, and an increased ability to answer questions about the data presented. These VR participants also more fully utilized the space provided in the virtual environment, moving objects further away from themselves, and placing them with less concern for their position relative to the real world. Some of these interaction differences may have stemmed from the different fields of view between the two HMDs.


\bibliographystyle{abbrv-doi}

\bibliography{template}
\end{document}